\documentclass[aps,pre,twocolumn,showpacs,notitlepage,floatfix,superscriptaddress]{revtex4-1}

\usepackage{amsfonts}
\usepackage{amsmath}
\usepackage{graphicx}
\usepackage{xcolor}
\usepackage{physics}
\usepackage{float}
\usepackage[colorlinks=true,linkcolor=blue,citecolor=blue,filecolor=blue,urlcolor=blue,pdfstartview=FitH,breaklinks=true]{hyperref}
\usepackage[sort&compress]{natbib}

\newcommand{\pksadd}{Max Planck Institute for the Physics of Complex Systems, Dresden D-01187, Germany}
\newcommand{\pcsadd}{Center for Theoretical Physics of Complex Systems, Institute for Basic Science(IBS), Daejeon 34126, Korea}
\newcommand{\ustadd}{Basic Science Program(IBS School), Korea University of Science and Technology(UST), Daejeon 34113, Korea}

\newcommand{\mg}{\mathcal{G}}
\newcommand{\mf}{\mathcal{F}}

\begin{document}

\title{Compact breathers generator in one-dimensional nonlinear networks}

\author{Carlo Danieli}
\affiliation{\pksadd}
\affiliation{\pcsadd}

\author{Alexei Andreanov}
\affiliation{\pcsadd}
\affiliation{\ustadd}

\date{\today}

\begin{abstract}
    Nonlinear networks can host spatially compact time periodic solutions called \emph{compact breathers}. Such solutions can exist accidentally (i.e. for specific nonlinear strength values) or parametrically (i.e. for any nonlinear strength). In this work we introduce an efficient generator scheme for one-dimensional nonlinear lattices which support either types of compact breathers spanned over a given number $U$ of lattice's unit cells and any number of sites $\nu$ per cell - scheme which can be straightforwardly extended to higher dimensions. This scheme in particular allows to show the existence and explicitly construct examples of parametric compact breathers with inhomogeneous spatial profiles -- extending previous results which indicated that only homogeneous parametric compact breathers exist. We provide explicit $d=1$ lattices with different $\nu$ supporting compact breather solutions for $U=1,2$.

\end{abstract}

\maketitle

\section{Introduction}

A remarkable feature of nonlinear non-integrable lattices is the existence of time-periodic spatially localized (typically exponentially) excitations called breather solutions~\cite{ovchinikov1970localized,sievers1988intrinsic,flach1998discrete,flach2008discrete,flach2005qBreathers,flach2006qBreathers}. Although forming a set of zero measure, such solutions are typically linearly stable and they impact the chaotic dynamics of the system as a generic trajectory may spend long times in their neighborhoods in phase space~\cite{tsironis1996slow,rasmussen2000statistical,eleftheriou2003breathers,eleftheriou2005discrete,gershgorin2005renormalized,matsuyama2015multistage,zhang2016dynamical} -- phenomenon visualized experimentally in superfluids \cite{Ganshin2009energy}, optical fibers \cite{Solli2007optical} and arrays of waveguides~\cite{Eisenberg1998discrete,Lederer2008discrete}. From exponentially localized solutions, the question whether discrete breathers can have zero tail turning into strictly \emph{compact breather} solutions naturally emerged. Page found spatially compact breather solutions in the \emph{Fermi-Pasta-Ulam-Tsingou} system in the limit of non-analytic box interaction potential~\cite{page1990asymptotic}, while Rosenau \emph{et.al.} showed the existence of traveling solitary waves with compact support in the modified \emph{Korteweg-de Vries} model~\cite{rosenau1993compactons}. Other successful attempts  include compact solutions in one-dimensional lattices in the presence of non-local nonlinear terms~\cite{kevrekidis2002bright}; compact solutions in discrete nonlinear Klein–Gordon models~\cite{Kevrekidis2003discrete}; and compact traveling bullet excitations (as well as super-exponentially localized moving breathers) in nonlinear discrete time quantum walks~\cite{Vakulchyk2018almost}. 

One possible mean to trigger spatially compact excitations in translationally invariant lattices which has become vastly popular in the recent years is destructive interference. Indeed, destructive interference in crystalline linear lattices may yield compact localized eigenstates (CLS) which are macroscopically degenerate and they form a dispersionless (or flat) band in the Bloch spectrum~\cite{derzhko2015strongly,Leykam2018artificial,leykam2018perspective}. Dubbed as flatband those networks supporting dispersionless bands, the intense activity around them is motivated from the experimental feasibility of CLS in a variety of different platforms, from photonics~\cite{mukherjee2015observation,vicencio2015observation,weimann2016transport}, to microwaves~\cite{bellec2013tight,casteels2016probing}, exciton-polariton~\cite{masumoto2012exciton} and ultra cold atoms~\cite{taie2015coherent}, among others.  

Therefore it was no surprise that the introduction of local Kerr nonlinearity in notable flatband networks yielded diverse examples of compact breather solutions~\cite{johansson2015compactification,gligoric2016nonlinear,belicev2017localized,bastian2018controlled,johansson2019nonlinear,stojanovic2020localized}. Their existence have been partly explained by a continuation criterium from linear CLS of flatband networks to compact breathers introduced in Ref.~\onlinecite{danieli2018compact} for CLS whose nonzero amplitudes are all equal. Meanwhile, compact time-periodic solutions have been found in one-dimensional nonlinear mechanical lattice networks~\cite{perchikov2017flat,perchikov2020stabilty} and in dissipative coherent perfect absorbers~\cite{danieli2020casting}. Furthermore, it has been found that compact discrete breathers induce Fano resonances~\cite{ramachandran2018fano} (similarly to discrete breathers~\cite{flach2003fano,vicencio2007fano}), and their existence is linked to nonlinear caging in linear lattices supporting only flat bands~\cite{gligoric2018nonlinear,diliberto2019nonlinear,danieli2020nonlinear,*danieli2020quantum}.  
 
In this work, we introduce a systematic scheme to generate nonlinear lattices with any number of sites $\nu$ per unit cell  supporting compact breather solutions spanning any given number $U$ of unit cells. This scheme is inspired and is based on the recently proposed single particle flatband generators~\cite{maimaiti2017compact,maimaiti2019universal,maimaiti2021flatband} -- schemes following and generalising previously proposed ones~\cite{flach2014detangling,dias2015origami,morales2016simple,rontgen2018compact,toikka2018necessary}. Our generator addresses systems supporting either \emph{accidental compact breathers} -- i.e. compact breather solutions existing at specific values of the nonlinear strength only -- or \emph{parametric compact breathers} -- i.e. compact breather existing at any value of the nonlinear strength. In particular, we are able to broaden the latter class of parametric compact breathers beyond the continuation criteria introduced in Ref.~\onlinecite{danieli2018compact} which rely on the spatial homogeneity of the compact solutions. Throughout the work, we provide explicit lattice samples with different number of bands supporting compact breather solutions for the cases $U=1$ and $U=2$ number of lattice unit cells. 

\section{Set-up and fundamental concepts}

Let us consider a one-dimensional nonlinear lattice with $\nu$ bands and nearest-neighbor hopping, whose equations of motion read
\begin{equation}
    i \dot{\Psi}_n = - H_0\Psi_n - H_1\Psi_{n+1} - H_1^\dagger\Psi_{n-1} + \gamma\mg(\Psi_n).
    \label{eq:FB_ham_NL1}
\end{equation}
For any $n\in\mathbb{Z}$, each entry of the time-dependent vector $\Psi_n = (\psi_n^1,\dots,\psi_n^\nu)^T\in\mathbb{C}^\nu$ represents one site of the lattice -- hence $\Psi_n$ represents its unit cell. The square matrices $H_0,H_1$ of size $\nu$ with Hermitian $H_0$, define the lattice profile, while $\mg$ is the nonlinear function -- chosen here such that $\mg(0) = 0$. We seek for lattices Eq.~\eqref{eq:FB_ham_NL1} which posses compact discrete breather solutions (CB), {\it e.g.} time-periodic spatially compact solutions 
\begin{align}
    C_{n,n_0} (t) &= \left[ \sum_{l=1}^{U} \Phi_l \delta_{n,n_0 + l}\right] e^{-i \Omega t}
    \label{eq:NL_FB_states1}
\end{align}
spanning over $U$ unit cells and with frequency $\Omega$. For convenience we refer to such breathers as breathers of size $U$. The vectors $\Phi_l = (\phi_l^1,\dots,\phi_l^\nu)^T$ for $1\leq l\leq U$ define the CB spatial profile. 

The ansatz Eq.~\eqref{eq:NL_FB_states1} is a solution of Eq.~\eqref{eq:FB_ham_NL1} if for all $1\leq j\leq U$
\begin{align}
    \label{eq:ex_cond_sol}
    & \Omega \Phi_j = -H_0\Phi_j - H_1\Phi_{j+1} - H_1^\dagger\Phi_{j-1} + \gamma\mg(\Phi_j),\\
    & \quad H_1 \Phi_1 =0 \qquad \qquad H_1^\dagger \Phi_U = 0 
    \label{eq:ex_cond_DI}
\end{align}
The conditions in Eq.~\eqref{eq:ex_cond_DI} ensure \emph{destructive interference} at the boundary of the compact sub-region occupied by the breather -- similarly to the flatband case discussed in Refs.~\onlinecite{maimaiti2017compact,maimaiti2019universal,maimaiti2021flatband} for linear compact localized eigenstates. For nonlinear lattices of Eq.~\eqref{eq:FB_ham_NL1} defined for given $H_0,H_1$, the classification of compact breathers is twofold: We distinguish CB based on the homogeneity of their spatial profiles on one hand, and their dependence on the nonlinearity strength $\gamma$ on the other hand. Specifically, a compact breather in Eq.~\eqref{eq:NL_FB_states1} is:
\begin{itemize}
    \item[(i)] 
    \emph{homogeneous}: if 
    \begin{equation}
        |\phi_l^j|^2 \in \{0,A_l^2\}\ , \quad 1\leq l\leq U\ ,\ \ 1\leq j\leq \nu
        \label{eq:CLS_H}
    \end{equation} 
    for a real $A\neq 0$ amplitude, while a CB is \emph{heterogeneous} otherwise.
    \item[(ii)] 
    \emph{accidental}: if it exists at specific fine-tuned values of the nonlinearity strength $\gamma_*\neq 0$. Otherwise, CB is \emph{parametric} if it exists for any value of the nonlinearity strength $\gamma$.
\end{itemize}
To clarify the context of these definitions, it's worth reminding, that breathers always come in families, e.g. if present they would exist for any strength of nonlinearity~\cite{flach1998discrete,flach2008discrete}, however they need not be compact always. So an accidental compact breather is part of family of breathers, that turn compact only for specific (isolated) control parameters. Consequently a parametric compact breather corresponds to a family of breathers that are compact for any parameter value.

\begin{figure}%[h]
    \centering
    \includegraphics[width=\columnwidth]{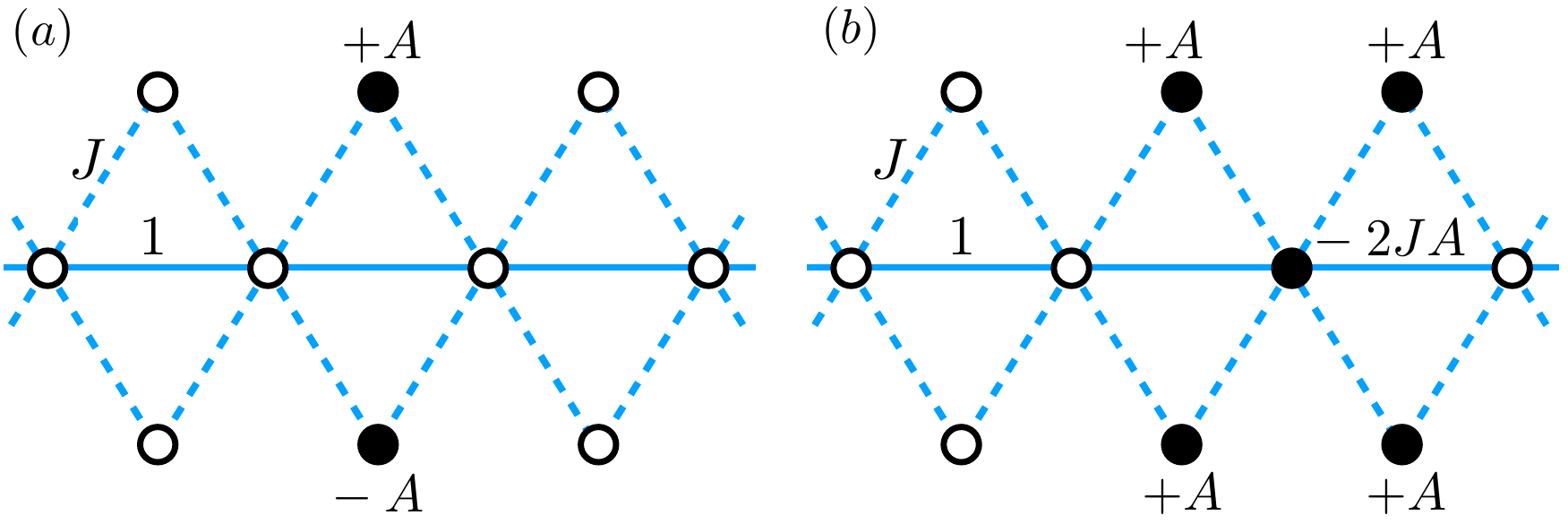}
    \caption{Example of $\nu=3$ network supporting both 
    % Eq.~\eqref{eq:diam_matr}, 
parametric compact breather - shown in (a) - and accidental compact breather - shown in (b). Both panels: hopping $1$ (solid lines) and $J$ (dashed lines). The black dots indicate the non-zero amplitudes of the breathers.}
    \label{fig:CB_samples}
\end{figure}

Following previous studies, it is known that homogeneous CB are parametric and they can be derived as a continuation of linear CLS of a flatband network into the nonlinear regime~\cite{danieli2018compact}. Otherwise, heterogeneous CB are instead typically accidental~\cite{johansson2015compactification}. We illustrate these distinctions with an example of a $\nu=3$ nonlinear network shown in Fig.~\ref{fig:CB_samples} with $\Psi_n = (a_n,b_n,c_n)$ and matrices 
\begin{equation}
    H_0 =  \begin{pmatrix}
        0 & J & 0 \\[0.3em]
        J & 0 & J \\[0.3em]
        0 & J & 0 
    \end{pmatrix},
    \quad
    H_1 =  \begin{pmatrix}
        0 & J & 0 \\[0.3em]
        0 & 1 & 0 \\[0.3em]
        0 & J & 0 
    \end{pmatrix},
    \label{eq:diam_matr}
\end{equation} 
in Eq.~\eqref{eq:FB_ham_NL1}, with $J>0$ the hopping parameter. In this example, we consider the local cubic nonlinear term $\mg(\Psi_n) = (|a_n|^2 a_n, |b_n|^2  b_n, |c_n|^2 c_n)$.

%    Figure 1
    
This lattice remarkably supports both types of the breathers -- parametric and accidental. Fig.~\ref{fig:CB_samples}(a)  shows a parametric $U=1$ compact breather present for any given $J\neq 0$ and any nonlinearity strength $\gamma\neq 0$, with frequency $\Omega = \gamma A^2$~\cite{danieli2018compact}. At the same time, Fig.~\ref{fig:CB_samples}(b) shows an accidental $U=2$ compact breather that at a given $J\neq \pm 1, \pm 1/2$~\footnote{For $J=1$, one finds $\gamma = 0$, while $J=\pm1/2$ implies $\gamma\rightarrow\infty$. However, in this latter case $J=\pm1/2$ the CB in Fig.~\ref{fig:CB_samples}(b) becomes homogeneous, but to exist it requires an additional onsite energy $-3/2$ on the central site.} it exists for the specific nonlinearity strengths  
\begin{equation}
    \gamma_* = 2 \frac{J^2 - 1}{4 J^2 - 1} 
    \label{eq:cb_cond1_ex}
\end{equation}
with frequency $\Omega_* = 2J^2 + \gamma A^2$. 

\section{Generating compact breathers}

We now outline schematically the generator scheme that we use to construct nonlinear lattices supporting compact breathers. This scheme is similar and is inspired by the generator scheme introduced in Refs.~\cite{maimaiti2017compact,maimaiti2019universal,maimaiti2021flatband} for linear flatband lattices: it relies on the same idea -- reconstruct, if possible, the nonlinear network from a given compact breather. The main steps for the construction of the network and the accidental compact breathers are
\begin{enumerate}
    \item Choose the nonlinear term $\mg$ in Eq.~\eqref{eq:FB_ham_NL1}
    \item Choose the desired size $U$ of the breather (in lattice unit cells) 
    \item Choose the nonlinearity strength $\gamma_*\neq 0$ and the breather frequency $\Omega_*$
    \item Fix the CB profile - $\{\Phi_l\}_{l=1}^U$ in Eq.~\eqref{eq:NL_FB_states1}
    \item Reconstruct the hopping matrices $H_0,H_1$ such that Eqs.~(\ref{eq:ex_cond_sol}-\ref{eq:ex_cond_DI}) support this CB
\end{enumerate} 
Parametric compact breathers can be generated from accidental breathers upon further fine-tuning by placing some suitable conditions on the generated hopping matrix $H_1$, as we show later. Let us also remark that for $U\geq 2$, the CB profile $\{\Phi_l\}_{l=1}^U$ is object to some nonlinear constraints that need to be resolved. This becomes obvious if one considers the last step in the above construction -- the reconstruction of the hopping matrices: assuming the profile $\Psi_l$, the frequency $\Omega$ and the nonlinearity is known, the problem of finding the matrices is a modification of an inverse eigenvalue problem presented in Ref.~\cite{boley1987survey}. It is then obvious that not any set of vectors $\Psi_l$  can be compatible with the desired block tridiagonal (or more generic, banded) structure of the Hamiltonian matrix of the linear problem.

In this work, we will focus for convenience on the case of local Kerr nonlinearity $\mg(\Psi_n) = \mf(\Psi_n) \Psi_n $  with
\begin{equation}
    \mf(\Psi_n) \equiv \sum_{j=1}^\nu |\psi_n^j|^{2\alpha} \  e_j\otimes  e_j
    \label{eq:FB_ham_NL1_2}
\end{equation}
However, the scheme can be applied for other types of nonlinearities $\mg$, including nonlocal ones. We first construct nonlinear networks supporting both accidental and parametric $U=1$ compact breathers. In this case, we also show the existence and provide explicit examples of parametric \emph{heterogeneous compact breathers}. Then, we  discuss the case of nonlinear networks supporting $U=2$ compact breathers. 

\subsection{Class $U=1$ breathers}
\label{sec:CBgen_U=1}

Let us construct a nonlinear network with $\nu$ sites per unit cell supporting accidental compact breather following the steps outlined above. We fix the frequency $\Omega_*$, the nonlinearity strength $\gamma_*$, and parameterise the amplitude vector as $\Phi_1 = \lambda_* \ket{n_1}$ with $\bra{n_1}\ket{n_1} = 1$, where $n_1$ is an arbitrary $\nu$ component vector. Then Eqs.~(\ref{eq:ex_cond_sol}-\ref{eq:ex_cond_DI}) reduce to
\begin{gather}
\label{eq:ex_cond_sol_U1}
    H_0 \ket{n_1} = - \Omega_* \ket{n_1} + \gamma_* \mf(\lambda_*\ket{n_1}) \ket{n_1}, \\
    H_1 \ket{n_1} = H_1^\dagger \ket{n_1} = 0. \notag
\end{gather}
From the r.h.s of Eq.~\eqref{eq:ex_cond_sol_U1} we define the vector 
\begin{gather}
    \ket{x_*} = -\Omega_* \ket{n_1} + \gamma_*\lambda_*^{2\alpha} \mf(n_1 ) \ket{n_1}.
    \label{eq:x_U1} 
\end{gather}
Assuming $\bra{x_*}\ket{n_1} \neq 0$ (for the case $\bra{x_*}\ket{n_1} = 0$ see Appendix~\ref{app:1}) and introducing a transverse projector $Q$ on $n_1$: $Q\ket{n_1}=0$, the matrices $H_0,H_1$ follow straightforwardly
\begin{gather}
    \label{eq:H0H1_U1}
    H_0 = \frac{\ket{x_*}\bra{x_*}}{\bra{x_*}\ket{n_1}} + Q_1 K Q_1\ ,\quad H_1 = Q_1 M Q_1
\end{gather}
where $K,M$ are arbitrary $\nu\times\nu$ matrices with $K$ Hermitian in order to ensure the Hermicity of $H_0$. The presence of this free parameters is expected, as every inverse eigenvalue problem has multiple solutions.
  
%    Figure 2
\begin{figure}%[h]
    \centering
    \includegraphics[width=0.85\columnwidth]{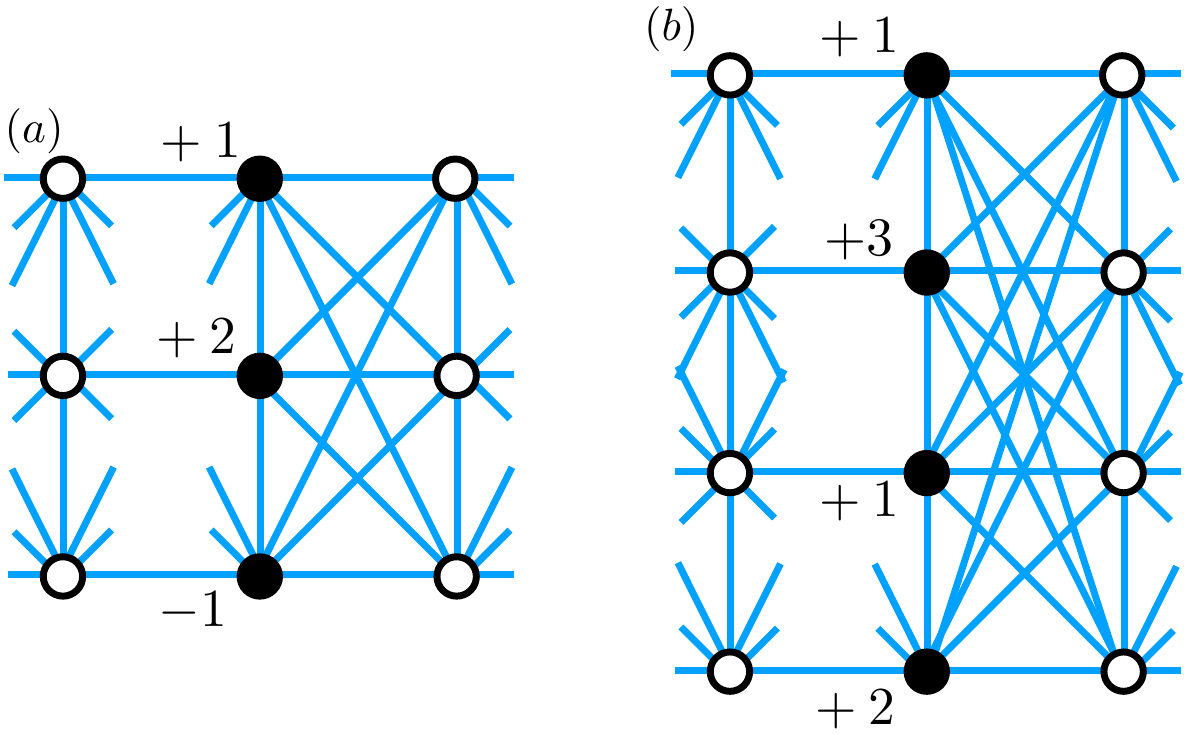}
    \caption{Examples of $U=1$ heterogeneous accidental compact breathers on $\nu=3$ network (a) and $\nu=4$ network (b). In both cases, $\lambda_*=1, \gamma_*=1$, and $\Omega_*=1$. The black dots indicate the non-zero amplitudes of the breathers.}
    \label{fig:CB_U1_nu3_4}
\end{figure}    

Examples of networks supporting $U=1$ accidental compact breathers are schematically shown in Fig.~\ref{fig:CB_U1_nu3_4} for $\nu=3$ [panel (a)] and $\nu=4$ [panel (b)]. The three components $\nu=3$ network reported in Fig.~\ref{fig:CB_U1_nu3_4}(a) has been constructed with $\ket{n_1} = \frac{1}{\sqrt{5}}(1,2,-1)$, while the four components $\nu=4$ network shown in Fig.~\ref{fig:CB_U1_nu3_4}(b) has been constructed with $\ket{n_1} = \frac{1}{\sqrt{15}}(1,3,1,2)$. In both cases, we chose $\lambda_*=1, \gamma_*=1$, and $\Omega_*=1$. Their correspondent matrices $H_0,H_1$ have been generated via Eq.~\eqref{eq:H0H1_U1}, and are presented in Appendix~\ref{app:CB_U1}.

Does the the nonlinear system Eq.~\eqref{eq:FB_ham_NL1} defined for fixed $H_0,H_1$ in Eq.~\eqref{eq:H0H1_U1} supports also \emph{parametric compact breathers} in addition to the accidental one at $\gamma = \gamma_*$? The answer depends on the number of the zero modes of $H_1$ {\it i.e.} whether $\ket{n_1}$ is the only zero mode of $H_1$ or not. 

If $\ket{n_1}$ is the only zero mode of $H_1$, then parametric compact breathers can exist if and only if the zero mode $\ket{n_1}$ of $H_1$ is homogeneous, {\it e.g.} has all of its components equal either to zero or to some real number $A$ -- as detailed in Appendix~\ref{app:1}. This case satisfies the criterium discussed in Ref.~\cite{danieli2018compact} and parametric compact breathers exists for any nonlinearity strength $\gamma$ with the profile $\Phi_1 = \lambda \ket{n_1}$ defined for any $\lambda$. The frequency $\Omega$ depends on $\gamma$ as
\begin{equation}
    \Omega = \Omega_* +  \gamma \frac{ (\lambda A)^{2\alpha}}{R^{2\alpha}} \left[ 1 - \frac{\gamma_*}{\gamma } \frac{\lambda_*^{2\alpha}}{ \lambda^{2\alpha}}\right] 
    \label{eq:CB_U1_freq}
\end{equation}
Here $A$ is the amplitude in every non-zero site of $\ket{n_1}$, and $R$ is the renormalization coefficient of $\ket{n_1}$ -- see Appendix~\ref{app:U1_sm} for details. 

If instead $H_1$ has multiple zero-modes, $\ket{n_l}$ (that can always be taken orthogonal $\bra{n_i}\ket{n_j} = \delta_{ij}$) the lattice defined for fixed $H_0,H_1$ in Eq.~\eqref{eq:H0H1_U1} can feature parametric compact breathers whose spatial profile does not require any homogeneity condition Eq.~\eqref{eq:CLS_H}~\cite{danieli2018compact}. We demonstrate this for the simplest case of two zero-modes $\ket{n_1}, \ket{n_2}$ and fixed $H_0,H_1$ obtained for given $\ket{n_1}, \gamma_*,\lambda_*, \Omega_*$ from Eq.~\eqref{eq:H0H1_U1}. We then search for a parametric CB solutions in the subspace of profiles $\Phi_1$ parameterised by $p, \lambda$:
\begin{equation}
    \Phi_1 = \lambda (\ket{n_1} + p\ket{n_2}) 
    \label{eq:CB_par_U1}
\end{equation}
By construction, for $\lambda = \lambda_*$ and $p=0$ in Eq.~\eqref{eq:CB_par_U1}, the network defined by the matrices $H_0,H_1$ in Eq.~\eqref{eq:H0H1_U1} supports an accidental compact breather at strength $\gamma_*$ and frequency $\Omega_*$. 

For $\gamma\neq \gamma_*$ and $\lambda \neq \lambda_*$ we search for $p\neq 0$ such that Eq.~\eqref{eq:CB_par_U1} is a CB solution of the network defined by the matrices $H_0,H_1$, Eq.~\eqref{eq:H0H1_U1}. The idea is to search a solution of Eq.~\eqref{eq:ex_cond_sol_U1} using ansatz~\eqref{eq:CB_par_U1} in terms of the unknown $\lambda$ and $p$. This results into a system of algebraic equations for $\lambda, p$. As detailed in Appendix~\ref{app:U1_mm}, we found that the real parameter $p$ can always be continuously expressed as a function of $(\gamma,\lambda)$, with $p(\gamma\to \gamma_*,\lambda \to \lambda_*) \to 0$. The resulting $\Phi_1(\gamma,\lambda, p(\gamma, \lambda))$ given by  Eq.~\eqref{eq:CB_par_U1} is a compact breather solution of the lattice defined by the hopping matrices $H_0,H_1$ for any $\gamma,\lambda$. Importantly there is no reason for the $\Phi_1$ constructed this way to be spatially homogeneous in general as we illustrate below with an example.  

%    Figure 3
\begin{figure}%[h]
    \centering
    \includegraphics [width=\columnwidth]{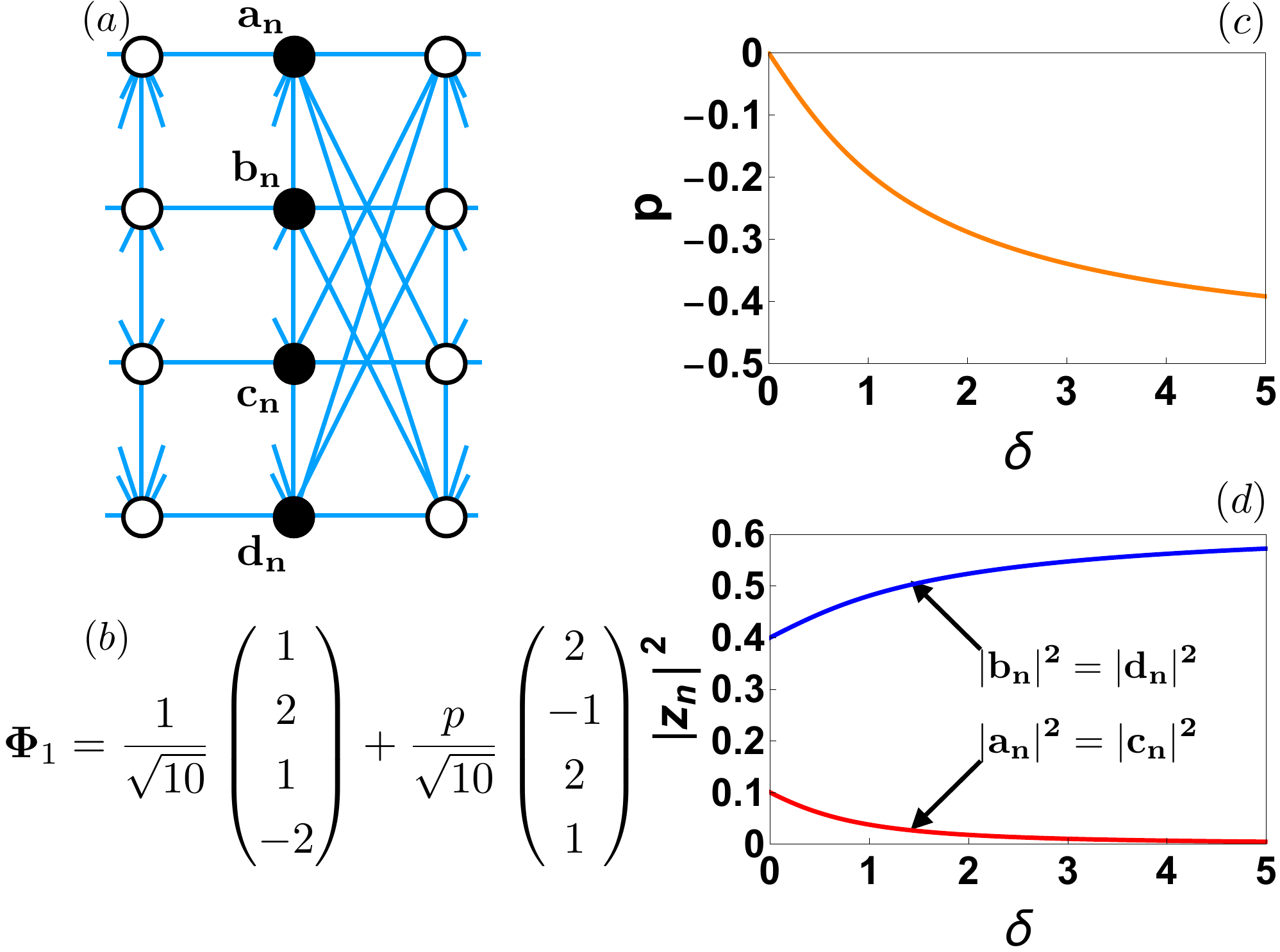}
    \caption{(a) Examples of $\nu=4$ networks which supports parametric heterogeneous compact breather. (b) parametrization of the breather spatial profile. (c) Root $p$ versus $\delta$, with $\delta$ control parameter $\gamma = \gamma_*+\delta$. (d) $|a_n|^2 = |c_n|^2$ (red) and $|b_n|^2 = |d_n|^2$ (blue) versus $\delta$. Here $\lambda_*=1, \gamma_*=1$, and $\Omega_*=1$. The black dots indicate the non-zero amplitudes of the breathers.}
    \label{fig:CB_U1_het}
\end{figure}

We illustrate this construction with an example shown in Fig.~\ref{fig:CB_U1_het} showing parametric heterogeneous compact breathers obtained via the above construction. We fix $\lambda = \lambda_* = 1$ in Eq.~\eqref{eq:CB_par_U1} and only vary $\gamma = \gamma_* + \delta$, where parameter $\delta$ controls the deviation away from $\gamma_*$. In panel~\eqref{fig:CB_U1_het}(a) we show the $\nu=4$ lattice with black dots indicating the nonzero amplitudes of the breathers. We choose $\Omega_*=1$, $\gamma_*=1$, $\ket{n_1} = \frac{1}{\sqrt{10}} (1,2,1,-2)$, $\ket{n_2} = \frac{1}{\sqrt{10}} (2,-1,2,1)$ and generate the hopping matrices $H_0,H_1$ via Eq.~\eqref{eq:H0H1_U1} supporting $\ket{n_1}$ as an accidental compact breather, while using the freedom in the choice of $H_1$, {\it e.g.} the appropriate choice of matrix $M$ in Eq.~\eqref{eq:H0H1_U1}, to ensure that $\ket{n_2}$ is the second zero mode of $H_1$ orthogonal to $\ket{n_1}$. Then the vector $\Phi_1$ is parameterised as $\Phi_1=\ket{n_1} + p \ket{n_2} = (a_n, b_n, c_n, d_n)$ -- where $n$ is the unit cell indexes; also see Eq.~\eqref{eq:CB_par_U1} -- as shown in Fig.~\ref{fig:CB_U1_het}(b). For the parametric compact breathers the parameter $p$ is expressed as a function of $\delta$ such that $p(\delta\to 0) \to 0$, as we discussed above. Figure~\eqref{fig:CB_U1_het}(d) shows the values of the components $|a_n|^2 = |c_n|^2$ (red curve) and $|b_n|^2 = |d_n|^2$ (blue curve) as a function of $\delta$: the difference between the values of $a_n$ and $c_n$ confirms that these parametric compact breathers are indeed heterogeneous. The details of this derivation are reported in Appendix~\ref{app:CB_het_fam}.

\subsection{Class $U=2$ breathers}
\label{sec:CBgen_U=2}

The previously discussed construction scheme directly extends to other compact breathers of larger sizes $U\geq 2$ albeit with some additional complications. In this section, we explicitly discuss the case of $U=2$ compact breathers and build explicit lattice network examples, but similar derivations can be performed for other values of $U$. 

Let us fix the number of bands $\nu$ and $\Omega_*, \gamma_*$. Then Eqs.~(\ref{eq:ex_cond_sol},\ref{eq:ex_cond_DI}) become for $U=2$
\begin{align}
    \label{eq:ex_cond_sol_U2_a}
    H_1\Phi_2 & = - (H_0 + \Omega_*) \Phi_1 + \gamma_*  \mf(\Phi_1) \Phi_1, \\
    \label{eq:ex_cond_sol_U2_b}    
    H_1^\dagger\Phi_1 & = - (H_0 + \Omega_*) \Phi_2 + \gamma_* \mf(\Phi_2) \Phi_2, \\
    H_1 \Phi_1 & = H_1^\dagger \Phi_2 = 0. \notag
\end{align}
Just like in the flatband case~\cite{maimaiti2017compact,maimaiti2019universal}, this system of equations can be regarded as an inverse eigenvalue problem for $H_1$ given $\Phi_1,\Phi_2$, while $H_0$ can be considered as a free parameter. However it is also simple to show that not every profile $\{\Phi_1, \Phi_2\}$ produces an $H_1$ -- the parametrization vectors $\{\Phi_1,\Phi_2\}$ are subject to nonlinear constraints, that ensure the existence of a solution $H_1$. One way to see that is notice that $\mel{\Phi_1}{H_1}{\Phi_2}$ can be computed independently from each of the above two equations~(\ref{eq:ex_cond_sol_U2_a}-\ref{eq:ex_cond_sol_U2_b}). The presence of such nonlinear constraints is a generic feature for any $U>1$. In our $U=2$ case we can pick independently $\Phi_1$ (alternatively $\Phi_2$), then the above equations yield polynomial constraints on the second vector $\Phi_2$ (alternatively $\Phi_1$) - see Appendix~\ref{app:2} for details of the derivation and resolution of the constraints. Assuming that these constraints are resolved and parameterizing the vectors as $\Phi_l = \lambda_{l*} \ket{n_l}$ with $\bra{n_l}\ket{n_l} = 1$ for $l=1,2$, we can use a simple ansatz for the hopping matrix $H_1 = \ket{u_*}\bra{v_*}$. Plugging it into Eqs.~(\ref{eq:ex_cond_sol_U2_a}-\ref{eq:ex_cond_sol_U2_b}) we find (see Appendix~\ref{app:2} for details)
\begin{align}
    \label{eq:U2_H1_u}
    \ket{u_*} & = - \lambda_{1*} (\Omega_* +  H_0) \ket{n_1} + \gamma_* \lambda_{1*}^{2\alpha+1} \mf( n_1) \ket{n_1}, \\
    \bra{v_*} & = - \lambda_{2*} \bra{n_2} (\Omega_* +  H_0) + \gamma_* \lambda_{2*}^{2\alpha+1} \bra{n_2} \mf(n_2).
    \label{eq:U2_H1_v}
\end{align}
By construction, $H_1$ satisfies Eq.~\eqref{eq:ex_cond_DI} - namely $H_1 \Phi_1 =0$ and $H_1^\dagger \Phi_U =0$. 

%    Figure 4
\begin{figure}%[h]
    \centering
    \includegraphics[width=0.55\columnwidth]{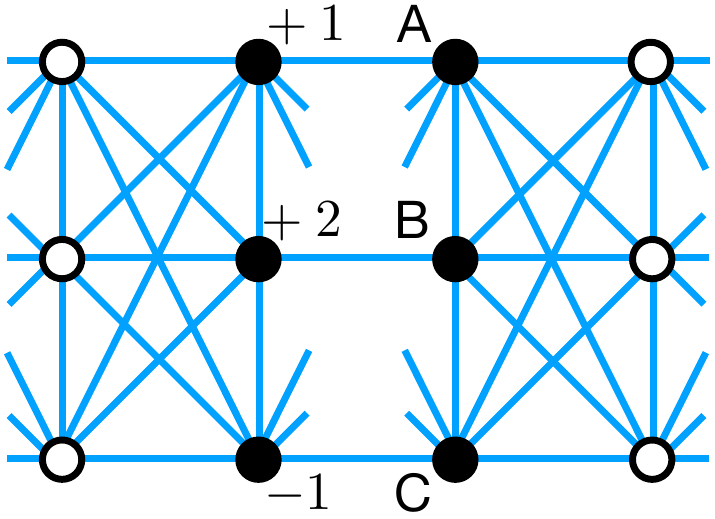}
    \caption{Example of $U=2$ heterogeneous accidental compact breathers on $\nu=3$ network. See the main text for sample amplitudes $A,B,C$. In this case $\lambda_{1*}= \lambda_{2*} =1$, $\Omega_*=1$, and $\gamma_*=1$. The black dots indicate the non-zero amplitudes.}
    \label{fig:CB_U2_nu3}
\end{figure}    

In Fig.~\ref{fig:CB_U2_nu3} we show samples of accidental $U=2$ compact breathers for $\nu=3$ network constructed following the above algorithm with $\ket{n_1} = \frac{1}{\sqrt{5}}(1,2,-1)$. We chose $\lambda_{1*} = \lambda_{2*} = \Omega_* = \gamma_* = 1$. By picking an arbitrary hermitian $H_0$ and resolving the nonlinear constraints the second parametrizing vector $\Phi_2 = (A,B,C)$ follows. For instance, two different choices of $H_0$ reported in Appendix~\ref{app:U2_ex} give two distinct vectors $\Psi_2$: $(A,B,C) =  (-1.464,  0.56026, -1.982)$ and $(A,B,C) = (-2.1328,  2.3334, -1.2823)$. Then we construct $H_1$ via Eqs.~(\ref{eq:U2_H1_u},\ref{eq:U2_H1_v}), defining the nonlinear network -- see Appendix~\ref{app:U2_ex} for details. 

The construction of heterogeneous parametric compact breathers for $H_1$ with multiple orthogonal zero-modes follows  the blueprint of the $U=1$ case discussed above. As a very cumbersome and involved procedure, we omit its presentation in the manuscript.

\section{Discussions and Perspectives}

In this work we have proposed a generator scheme for one-dimensional nonlinear lattices supporting discrete compact breathers. This scheme follows the generator schemes recently proposed in Refs.~\onlinecite{maimaiti2017compact,maimaiti2019universal,maimaiti2021flatband} for the single particle flatband networks, and we we have explicitly applied our results to the case of local Kerr nonlinearity, In particular, we have outlined explicitly the generator for compact breathers spanning over $U=1$ and $U=2$ unit cells, and we have presented several example nonlinear lattices supporting such solutions. Furthermore, we have successfully shown the existence and explicitly constructed nonlinear lattices supporting parametric heterogeneous compact breathers. These solutions substantially widen the class of parametric compact breathers, as the formerly known families of compact breathers follow as continuation in the nonlinear regime of spatially homogeneous linear CLS of flatband networks according to the criterium discussed in Ref.~\onlinecite{danieli2018compact}. 

The proposed compact breather generator for one-dimensional nonlinear networks and applied for the case of local Kerr nonlinearity acts as a blueprint and it can be employed for other types of nonlinear contributions $\mg$ in Eq.~\eqref{eq:FB_ham_NL1} - \textit{e.g.} saturable, nonlocal, among others, as well as in higher-dimensional nonlinear networks following the flatband generator discussed in Ref.~\onlinecite{maimaiti2021flatband}. This scheme is the first one addressing classical interacting systems, complementing those recently proposed in Refs.~\onlinecite{santos2020methods,danieli2020many} for quantum many-body systems featuring localization properties. Our results allow to devise nonlinear lattices capable to support exact compactly localized excitations based on the principle of destructive interference, broadening the successful research areas of localization of linear lattices due to flatbands onto novel nonlinear structures.  

\begin{acknowledgments}
    The authors thank Sergej Flach for helpful discussions. This work was supported by the Institute for Basic Science, Korea (IBS-R024-D1).
\end{acknowledgments}

\appendix

\section{class $U=1$ compact breathers}
\label{app:1}

The matrices $H_0,H_1$ in Eqs.~(\ref{eq:x_U1},\ref{eq:H0H1_U1}) 
\begin{align}
    \label{eq:H0H1_U1_app}
    & H_0 = \frac{\ket{x_*}\bra{x_*}}{\bra{x_*}\ket{n_1}} + Q_1 K Q_1\ ,\quad 
    H_1 = Q_1 M Q_1\\
    \label{eq:x_U1_app} 
    & \quad \ket{x_*} = -\Omega_* \ket{n_1} + \gamma_*\lambda_*^{2\alpha} \mf(\ket{n_1}) \ket{n_1}
\end{align}
supporting compact breathers are obtained for fixed $\Omega_*, \gamma_*$ under the assumption that $\bra{x_*}\ket{n_1} \neq 0$. In the case $\bra{x_*}\ket{n_1} = 0$, $H_0$ turns to
\begin{gather}
    \label{eq:H0_U1_alt_app}
    H_0 = \frac{\ket{x_*}\bra{n_1} + \ket{n_1}\bra{x_*}}{\bra{n_1}\ket{n_1}} + Q_1 K Q_1
\end{gather}
Let us search whether they support breather for generic frequency $\Omega$ and nonlinear strength $\gamma$ according to the number off zero modes for $H_1$. 

\subsection{Single zero-mode of $H_1$}
\label{app:U1_sm}

Let us parametrize $\Phi_1 = \lambda \ket{n_1}$ supposing $H_1$ in Eq.~\eqref{eq:H0H1_U1_app} with one single zero-mode $\ket{n_1}$. Eq.~\eqref{eq:x_U1_app}  for generic $\Omega$ and $\gamma$ turns to
\begin{align}
    \label{eq:U1_pop_app}
    & \ket{x} = - \Omega \ket{n_1} + \gamma\lambda^{2\alpha} \mf(n_1) \ket{n_1} 
\end{align}
From Eq.~\eqref{eq:ex_cond_sol_U1} -- here recalled $\Omega \Phi_1 = [- H_0 +  \gamma  \mf( \Phi_1)] \Phi_1$ -- the identity follows
\begin{align}
    \label{eq:U1_pop2_app}
    & \left[ \gamma \lambda^{2\alpha} -  \gamma_*\lambda_*^{2\alpha}\right]  \mf(n_1) \ket{n_1} = (\Omega - \Omega_*)\ket{n_1}
\end{align}
If $\ket{n_1}$ is homogeneous Eq.~\eqref{eq:CLS_H} \textit{i.e.} $\ket{n_1} = (z_1,\dots,z_\nu)/R$ where all complex entrees are either zero or $|z_j|^2= A^2$ with $R$ the renormalization coefficient, it follows that 
\begin{gather}
    \label{eq:U1_pop2b_app}
    \mf( n_1  ) \ket{n_1} = \frac{A^{2\alpha}}{R^{2\alpha}} \ket{n_1}
\end{gather}
leading to the parametrization of the frequency $\Omega$ 
\begin{gather}
    \label{eq:U1_pop4_app}
    \Omega = \Omega_* +  \gamma \frac{ (\lambda A)^{2\alpha}}{R^{2\alpha}} \left[ 1 - \frac{\gamma_*}{\gamma } \frac{\lambda_*^{2\alpha}}{ \lambda^{2\alpha}} \right] 
\end{gather}
If instead $\ket{n_1}$ is not homogeneous then Eq.~\eqref{eq:U1_pop2b_app} fails. For any $z_j\neq 0$ in $ \ket{n_1}$ Eq.~\eqref{eq:U1_pop2_app} turns to 
\begin{eqnarray}
    \label{eq:U1_pop3c_app}
    &  \left[ \gamma \lambda^{2\alpha} -  \gamma_*\lambda_*^{2\alpha}\right]  |z_j|^{2\alpha} = R^{2\alpha} (\Omega - \Omega_*) 
\end{eqnarray}
which, since $|z_j|^{2\alpha} \neq |z_i|^{2\alpha}$, it  is solved when both sides are zero, {\it i.e.} 
\begin{eqnarray}
    \label{eq:U1_pop5_app}
    &  \gamma  =   \gamma_* \left(\frac{ \lambda_*}{\lambda} \right)^{2\alpha} \quad\qquad \Omega = \Omega_*
\end{eqnarray}
corresponding to a trivial solution with fixed frequency $\Omega$ and rescaling the nonlinearity $\gamma$ corresponds to rescale the prefactor of the breather. 

\subsubsection{Examples}
\label{app:CB_U1}

{\it Example 1}:
let us consider $\nu=3$ band networks, and $\ket{n_1} = \frac{1}{\sqrt{5}}(1,2,-1)$ - shown in Fig.~\ref{fig:CB_U1_nu3_4}(a). Eq.~\eqref{eq:H0H1_U1_app} for $K=\mathbb{O}_3$ and $M=\mathbb{I}_3$ yields
\begin{equation}
 \footnotesize
\begin{split}
    H_0 &= 
    \begin{pmatrix}
        -0.231481 & -0.185185 & 0.231481 \\[0.3em]
        -0.185185 & -0.148148 & 0.185185 \\[0.3em]
        0.231481& 0.185185 & -0.231481 
    \end{pmatrix}\\
    &\qquad \qquad H_1 = \frac{1}{6}
    \begin{pmatrix}
        5 &  -2 & 1\\[0.3em]
        -2 &  2 & 2 \\[0.3em]
        1 & 2 & 5
    \end{pmatrix}
\end{split}
    \label{eq:cb_nu3_ex1}
\end{equation}
while for $K,M$ random matrixes with integer entrees between $4$ and $-4$ (with $K$ hermitian)
\begin{equation}
 \footnotesize
\begin{split}
    H_0 &= 
    \begin{pmatrix}
        -2.56481 & 3.23148 & 4.73148 \\[0.3em]
        3.23148 & -1.14815 & 1.60185  \\[0.3em]
        4.73148 & 1.60185 & 7.10185 
    \end{pmatrix}\\
    &\qquad \quad H_1 = \frac{1}{3}
    \begin{pmatrix}
        17 &  8 & 33 \\[0.3em]
        2 &  0 & 2 \\[0.3em]
        21 & 8 & 37
    \end{pmatrix}
\end{split}
    \label{eq:cb_nu3_ex1b}
\end{equation}

\noindent
{\it Example 2}:
let us consider $\nu=4$ band networks, and $\ket{n_1} = \frac{1}{\sqrt{15}}(1,3,1,2)$ - shown in Fig.~\ref{fig:CB_U1_nu3_4}(b). Eq.~\eqref{eq:H0H1_U1_app} for $K=\mathbb{O}_4$ and $M=\mathbb{I}_4$ yields
\begin{equation}
 \footnotesize
\begin{split}
    H_0 &= - 
    \begin{pmatrix}
        0.103704 &0.133333 &0.103704 &0.162963  \\[0.3em]
        0.133333 &0.171429 &0.133333 &0.209524  \\[0.3em]
        0.103704 &0.133333 &0.103704 &0.162963  \\[0.3em]
        0.162963 &0.209524 &0.162963 &0.256085 
    \end{pmatrix}\\
    &\qquad \qquad H_1 = \frac{1}{15}
    \begin{pmatrix}
        14 &  -3 & - 15 & -2 \\[0.3em]
        -3 &  6 & -3 & -6 \\[0.3em]
        -1 & -3 & 14 & -2 \\[0.3em]
        -2 & -6 & -2 & 11
    \end{pmatrix}
\end{split}
    \label{eq:cb_nu4_ex1}
\end{equation}
Similarly to the previous examples, different matrixes $H_0,H_1$ can be obtained by including $K,M$ as random matrixes in Eq.~\eqref{eq:H0H1_U1_app}.

\subsection{Multiple zero-modes of $H_1$}
\label{app:U1_mm}

Let us suppose the matrix $H_1$ has two zero-modes $\ket{n_1}$ and $\ket{n_2}$, with $\bra{n_i}\ket{n_j} = \delta_{i,j}$. We insert their linear combination $\lambda \ket{n_1} + \mu \ket{n_2}$ in Eq.~\eqref{eq:ex_cond_sol_U1}, which read
\begin{equation}
	\begin{split}
	    &\quad \left[\lambda + \mu\frac{ \bra{x_*}\ket{n_2}}{ \bra{x_*}\ket{n_1}} \right] \ket{x_*} 
+ \lambda \Omega \ket{n_1} \\
	    & + \mu\left[ Q_1 K Q_1 + \Omega\right] \ket{n_2} 
		- \gamma \ket{\mg(\lambda n_1 + \mu n_2)} = 0
	\end{split}
    \label{eq:CB_U1_2m_1}
\end{equation}
For compactness, we keep the notation $\mg$ in Eq.~\eqref{eq:FB_ham_NL1_2}. 
Inserting $\ket{x_*}$ from Eq.~\eqref{eq:x_U1_app} in the above expression returns
\begin{equation}
	\begin{split}
		0 &= \left[\lambda (\Omega - \Omega_*) - \mu \Omega_* \frac{\bra{x_*}\ket{n_2}}{\bra{x_*}\ket{n_1}} \right]  
 		\ket{n_1} + \mu\left[ Q_1 K Q_1 + \Omega\right] \ket{n_2} \\
 		& - \gamma \ket{\mg(\lambda n_1 + \mu n_2)}
		+ \gamma_*\lambda_*^{2\alpha} \left[ \lambda + \mu \frac{\bra{x_*}\ket{n_2}}{\bra{x_*}\ket{n_1}} \right] \ket{\mg(n_1)} 
	\end{split}
    \label{eq:CB_U1_2m_2}
\end{equation}
We now project Eq.~\eqref{eq:CB_U1_2m_2} onto $\ket{n_1}$, $\ket{n_2}$. 
This yields two equations, namely
\begin{align}
    \label{eq:CB_U1_2m_3_n1}
   & \lambda (\Omega - \Omega_*)  - \mu \Omega_* \xi_{1,2}- \gamma \bra{n_1}\ket{\mg(\lambda n_1 + \mu n_2)} \\
        &\qquad\qquad + \gamma_*\lambda_*^{2\alpha} \left[ \lambda + \mu \xi_{1,2} \right] \bra{n_1}\ket{\mg(n_1)} = 0 \notag\\
    \label{eq:CB_U1_2m_3_n2}
        & \mu\left[ \mel{n_2}{Q_1 K Q_1}{n_2} + \Omega\right]  - \gamma \bra{n_2}\ket{\mg(\lambda n_1 + \mu n_2)} \\ 
        & \qquad\qquad  + \gamma_*\lambda_*^{2\alpha}  \left[ \lambda + \mu \xi_{1,2} \right] \bra{n_2}\ket{\mg(n_1)} = 0 \notag
\end{align}
where we introduced the coefficient
\begin{gather}
    \xi_{1,2} = \frac{\bra{x_*}\ket{n_2}}{\bra{x_*}\ket{n_1}}
    \label{eq:cb_u1_c}
\end{gather}
We parametrize the vector $\Phi_1$ for the parameter $p,\lambda$
\begin{equation}
    \Phi_1 = \lambda (\ket{n_1} + p \ket{n_2})
    \label{eq:CB_par_U1_app}
\end{equation} 
Eq.~\eqref{eq:CB_U1_2m_3_n1} yields the frequency $\Omega$ of the parametric compact breather, which is parametrically dependent on the parameter $p$
\begin{align}
    \label{eq:CB_U1_2m_omega}
        \Omega &=  \Omega_* \left[1 + p \xi_{1,2}\right] + \gamma \lambda ^{2\alpha } \bra{n_1}\ket{\mg( n_1 + p n_2)} \\
        & - \gamma_*\lambda_*^{2\alpha} \left[ 1 + p \xi_{1,2} \right] \bra{n_1}\ket{\mg(n_1)} \notag
\end{align}
This expression for $\Omega$ plugged in Eq.~\eqref{eq:CB_U1_2m_3_n2} yields the following polynomial - called $f_1$ -  in $p$ 
\begin{align}
    \label{eq:CB_U1_f1}
        f_1(p) & \equiv p \Omega_* \left[1+ p \xi_{1,2}\right] +  p \mel{n_2}{Q_1 K Q_1}{n_2} \\
        & +\gamma_*\lambda_*^{2\alpha} \left[ 1 + p  \xi_{1,2} \right] 
  \left[  \langle n_2  | \mathcal{G}(n_1)\rangle   - p  \langle n_1  | \mg(n_1)\rangle   \right]   \notag \\
        & + \gamma \lambda ^{2\alpha } \left[ p \bra{n_1}\ket{\mg( n_1 + p n_2)} - \bra{n_2}\ket{\mg( n_1 + p n_2)} \right]  \notag
\end{align}
For values of $\gamma\neq \gamma_*$ and $\lambda \neq \lambda_*$ violating the equality $ \gamma_*\lambda_*^{2\alpha}   =  \gamma \lambda^{2\alpha}$, 
we computed the roots $p(\gamma,\lambda)$ of $f_1$ in Eq.~\eqref{eq:CB_U1_f1} which are functions of $\gamma,\lambda$. We then track the root such that $p(\gamma\rightarrow \gamma_*,\lambda \rightarrow \lambda_*) \rightarrow 0$. We use this root to parametrizes a compact breather  in Eq.~\eqref{eq:CB_par_U1_app}. at certain values of $ \gamma$. The frequency $\Omega$ is provided by Eq.~\eqref{eq:CB_U1_2m_omega}. 

\subsubsection{Example}
\label{app:CB_het_fam}

%{\it Example 5}:
Let us consider $\nu=4$ band networks. We then consider $\ket{n_1} = \frac{1}{10}(1,2,1,-2)$, and via Eq.~\eqref{eq:H0H1_U1_app} with $M=\mathbb{I}_2$ we obtain the matrix $H_1$
\begin{equation}
 \footnotesize
    \begin{split}
        H_1 = \frac{1}{2}
        \begin{pmatrix}
            1 &  0 & - 1 & 0 \\[0.3em]
            0 &  1 & 0 & 1 \\[0.3em]
            -1 & 0 & 1 & 0 \\[0.3em]
            0 & 1 & 0 & 1
        \end{pmatrix}
    \end{split}
    \label{eq:cbhet_nu4_ex1_H1}
\end{equation}
We observe that $H_1$ has the vector $\ket{n_2} = \frac{1}{10}(2,-1,2,1)$ as zero more, with $\bra{n_1}\ket{n_2} = 0$. We construct $H_0$  via  Eq.~\eqref{eq:H0H1_U1_app} with $K=\mathbb{O}_4$, $\lambda_*=1$, $\gamma_*=1$ and $\Omega_*=1$
\begin{equation}
 \footnotesize
    H_0 = 
    \begin{pmatrix}
        -0.122727 & -0.163636& -0.122727& 0.163636  \\[0.3em]
        -0.163636 & -0.218182& -0.163636& 0.218182  \\[0.3em]
        -0.122727 & -0.163636& -0.122727& 0.163636  \\[0.3em]
        0.163636 & 0.218182 &0.163636& -0.218182
    \end{pmatrix}
    \label{eq:cbhet_nu4_ex1_H0}
\end{equation}
In this case, we keep $\lambda = \lambda_* = 1$ and vary only $\gamma$. The polynomial $f_1$ in Eq.~\eqref{eq:CB_U1_f1} reduces to
%(dividing by 0.12 
\begin{align}
\label{eq:f1_CB_ex1}
        f_1(p)& = -1 + 5.32 p +  p^2 \\
        & +  \gamma (1 + p (-1.167 + p (-6 + p (1.167 + p))))\notag
\end{align}
We observe that for $\gamma = \gamma_* = 1$, $f_1(p=0) = 0$. We focus on the root $p(\gamma)$ that for $\gamma\rightarrow \gamma_*$ converges to zero to parametrize $\Phi_1(\gamma) = \ket{n_1} + p(\gamma)\ket{n_2}$.

\section{Class $U=2$ accidental compact breathers}
\label{app:2}

Eqs.~(\ref{eq:ex_cond_sol_U2_a},\ref{eq:ex_cond_sol_U2_b}) and the destructive interference condition Eq.~\eqref{eq:ex_cond_DI} restated for the vectors $\Phi_l = \lambda_{l*} \ket{n_l}$ with $\bra{n_l}\ket{n_l} = 1$ for $l=1,2$ read
\begin{footnotesize}
\begin{align}
    \label{eq:ex_cond_sol_U2_app_a_2}
    & \lambda_{2*} H_1 \ket{n_2} = - \lambda_{1*} (\Omega_* + H_0) \ket{n_1} + \gamma_* \lambda_{1*}^{2\alpha + 1} \mf( n_1 ) \ket{n_1} \\
    \label{eq:ex_cond_sol_U2_app_b_2}
    & \lambda_{1*} \bra{n_1} H_1 = - \lambda_{2*}  \bra{n_2} (\Omega_* + H_0) + \gamma_*\lambda_{2*}^{2\alpha + 1} \bra{n_2} \mf( n_2 ) \\
    & \quad H_1\ket{n_1} = 0 \qquad \qquad H_1^\dagger \ket{n_2} = 0 
    \label{eq:ex_cond_sol_U2_app_c_2}
\end{align}
\end{footnotesize}
where the matrix $H_0$ is a free parameter of the problem.\\
\\
These equations give
\begin{align}
    \label{eq:U2_ids_app_a}
    \lambda_{1*}\lambda_{2*} \mel{n_1}{H_1}{n_2} = -& \lambda_{1*}^2 \mel{n_1}{\Omega_* + H_0}{n_1}  \\
     +\gamma_* &\lambda_{1*}^{2\alpha + 2} \mel{n_1}{\mf(n_1)}{n_1} \notag \\
    \label{eq:U2_ids_app_b}
   \lambda_{1*}\lambda_{2*} \mel{n_1}{H_1}{n_2} = -&\lambda_{2*}^2 \mel{n_2}{\Omega_* + H_0}{n_2} \\
     +\gamma_*& \lambda_{2*}^{2\alpha + 2} \mel{n_2}{\mf( n_2 )}{n_2} \notag\\
 \label{eq:U2_ids_app_c}   
   \Omega_* \bra{n_2}\ket{n_1} = - \mel{n_2}{H_0}{n_1}& + \gamma_* \lambda_{1*}^{2\alpha} \mel{n_2}{\mf( n_1 )}{n_1} \\ 
   \label{eq:U2_ids_app_d} 
 \Omega_* \bra{n_2}\ket{n_1} = - \mel{n_2}{H_0}{n_1}& + \gamma_* \lambda_{2*}^{2\alpha} \mel{n_2}{\mf( n_2 )}{n_1}  
\end{align} 
ultimately yielding the following identities
\begin{align}
\label{eq:U2_ids2_app_a_2}
&\quad \lambda_{2*}^2 \mel{n_2}{\Omega_* + H_0}{n_2} - \lambda_{1*}^2 \mel{n_1}{\Omega_* + H_0}{n_1} \\
&= \gamma_* \left[ \lambda_{2*}^{2\alpha + 2}  \mel{n_2}{\mf( n_2 )}{n_2} - \lambda_{1*}^{2\alpha + 2} \mel{n_1}{\mf( n_1 )}{n_1} \right] \notag \\
& \qquad  \lambda_{1*}^{2\alpha} \mel{n_2}{\mf( n_1 )}{n_1} = \lambda_{2*}^{2\alpha} \mel{n_2}{\mf( n_2 )}{n_1}
    \label{eq:U2_ids2_app_b_2}
\end{align} 
Let us fix {\it e.g.} $\ket{n_1}$. Then, Eq.~(\ref{eq:U2_ids_app_c},\ref{eq:U2_ids2_app_a_2},\ref{eq:U2_ids2_app_b_2}) form a system of polynomial equations which set the necessary conditions for $\ket{n_2}$.\\
\\
With $\ket{n_1},\ket{n_2}$ defined, we then generate $H_1$ of  the form $H_1 = \ket{u_*} \bra{v_*}$ where $\bra{v_*}\ket{n_1} = 0 = \bra{n_2}\ket{u_*}$  so Eq.~\eqref{eq:ex_cond_sol_U2_app_c_2} is met. In Eqs.~(\ref{eq:ex_cond_sol_U2_app_a_2},\ref{eq:ex_cond_sol_U2_app_b_2}) this yields
\begin{footnotesize}
\begin{align}
    \label{eq:U2_H1_ids_a}
        \ket{u_*} \bra{v_*}\ket{n_2} &= - (\Omega_* + H_0)\ket{n_1} + \gamma_* \lambda_{1*}^{2\alpha+1} \mf( n_1) \ket{n_1} \\
        \bra{n_1}\ket{u_*} \bra{v_*} &= -\bra{n_2}(\Omega_* +  H_0) + \gamma_* \lambda_{2*}^{2\alpha+1} \bra{n_2} \mf( n_2) 
    \label{eq:U2_H1_ids_b}
    \end{align} 
\end{footnotesize}    
to which follows the parametrization
\begin{align}
    \label{eq:U2_H1_u_app}
    \ket{u_*} &= - \lambda_{1*} (\Omega_* +  H_0) \ket{n_1} +  \gamma_* \lambda_{1*}^{2\alpha+1} \mf( n_1) \ket{n_1} \\
    \bra{v_*} &= - \lambda_{2*} \bra{n_2} (\Omega_* +  H_0) +  \gamma_* \lambda_{2*}^{2\alpha+1} \bra{n_2} \mf( n_2)   
    \label{eq:U2_H1_v_app}
\end{align} 

\subsubsection{Example}
\label{app:U2_ex}

Let us consider $\nu=3$ band networks, and $\ket{n_1} = \frac{1}{\sqrt{5}}(1,2,-1)$ - shown in Fig.~\ref{fig:CB_U2_nu3} parametrizing the vector $\Phi_1= \lambda_{1*} \ket{n_1}$. We fix $\lambda_{1*}=1=\lambda_{2*}$, the nonlinear strength $\gamma_{*}=1$ and the frequency $ \Omega_{*}=1$.

\noindent
{\it Example 1}:
let us choose the hermitian matrix $H_0$ as 
\begin{equation}
 \footnotesize
    \begin{split}
        H_0 &= \begin{pmatrix}
            0 & 0& 1 \\[0.3em]
            0 & 1 & 0 \\[0.3em]
            1& 0 & -2 
        \end{pmatrix}
    \end{split}
    \label{eq:cb_U2nu3_ex1_H0}
\end{equation}
The system of polynomial equations~(\ref{eq:U2_ids_app_c},\ref{eq:U2_ids2_app_a_2},\ref{eq:U2_ids2_app_b_2}) yields
\begin{equation}
    \begin{split}
        \Phi_2= (-1.464,  0.56026, -1.982)
    \end{split}
    \label{eq:cb_U2nu3_ex1_Phi2}
\end{equation}
and Eqs.~(\ref{eq:U2_H1_u_app},\ref{eq:U2_H1_v_app}) define the hopping matrix
\begin{equation}
 \footnotesize
    \begin{split}
        &  H_1 = \frac{1}{6}
        \begin{pmatrix}
            3.65595 & -0.175859 & 3.30423 \\[0.3em]
            -29.2476 & 1.40687 & -26.4338 \\[0.3em]
            -10.9678 & 0.527577& -9.91268
        \end{pmatrix}
    \end{split}
    \label{eq:cb_U2nu3_ex1_H1}
\end{equation}

\noindent
{\it Example 2}:
let us choose the hermitian matrix $H_0$ as 
\begin{equation}
 \footnotesize
    \begin{split}
        H_0 &= 
        \begin{pmatrix}
            4 & 8& 9 \\[0.3em]
            8 & 2 & 3 \\[0.3em]
            9& 3 & 6 
        \end{pmatrix}
    \end{split}
    \label{eq:cb_U2nu3_ex2_H0}
\end{equation}
The system of polynomial equations~(\ref{eq:U2_ids_app_c},\ref{eq:U2_ids2_app_a_2},\ref{eq:U2_ids2_app_b_2}) yields
\begin{equation}
    \begin{split}
        \Phi_2= (-2.1328,  2.3334, -1.2823)
        \end{split}
    \label{eq:cb_U2nu3_ex2_Phi2}
\end{equation}
and Eqs.~(\ref{eq:U2_H1_u_app},\ref{eq:U2_H1_v_app}) define the hopping matrix
\begin{equation}
 \footnotesize
    \begin{split}
        &H_1 = \frac{1}{6}
        \begin{pmatrix}
            -98.7015 & -64.5789 & -227.859\\[0.3em]
            -134.593 & -88.0622 & -310.717 \\[0.3em]
            -80.7558 & -52.8373 & -186.43
        \end{pmatrix}
    \end{split}
    \label{eq:cb_U2nu3_ex2_H1}
\end{equation}

\bibliography{general,flatband,mbl,ergodicity}

\end{document}